# A Graphical Approach for Friedman Test: Moments Approach

Elsayed A. H. Elamir[1]

Management & Marketing Department, College of Business,
University of Bahrain, P.O. Box 32038,
Kingdom of Bahrain

## Abstract

Friedman test is a nonparametric method that proposed for analyzing data from a randomized complete block design as a robust alternative to parametric method and widely applied in many fields such as agriculture, biology, business, education, and medicine. After the null hypothesis of no treatment effects is rejected, the post-hoc pairwise comparisons must be applied to identify where the differences occur. As the number of groups increases, the number of required comparisons becomes large and this may increase the type I error. The aim of this study is twofold. The main aim is to suggest expression that facilitates the plotting Friedman test by gathering the test and pairwise comparisons in one simple step. The second aim is to derive the sampling distribution of the suggested expression by utilizing method of moments that helps in obtaining the decision limit. An application and simulation study are carried out to show the advantage of the suggested method and to compute the empirical type I error. The results are of great value where the proposed method makes huge reduction in the number of required tests to show where the discrepancies occur, holds the type I error close to the nominal value and provides visual, deep insight and understanding where the treatment effects occur.

**Keywords:** Bonferroni approximation; chi square distribution; gamma distribution; multiple comparisons; nonparametric tests.

Running title: Graphical approach for Friedman test.

[1]Email: shabib@uob.edu.bh

# 1 Introduction

Friedman test is the nonparametric analogue of the usual randomized block analysis of variance (ANOVA) and was developed by Friedman (1937). If the underlying responses may not be assumed to follow a normal distribution or if only the ranks of the responses are available, Friedman test may be used to perform an appropriate analysis. It is assumed that *B* blocks (ie, sets of homogeneous experimental units each) are available. Within each block, each of the treatments is applied to one experimental unit; see, Hettmansperger (1984), Gibbons and Chakraorti (2003), Sheskin (2000) and Kandetody et al. (2021).

In recent years, there are interest and new applications of Friedman test in many fields such as agriculture, biology, bioinformatic, business and medicine. Abdel-Basset and Shawky (2019) used Friedman test to analyse the effect between flower pollination algorithm and six different metaheuristics such as "genetic algorithm", "cuckoo search", "grasshopper optimization algorithm", and others on a constrained engineering optimization problem. By using Friedman test, they indicated that flower pollination algorithm is better than other competitors in solving the given problem. Hou and Wang (2018) performed test on the accuracy of the frameworks that combines "improved feature extraction algorithm scale invariant feature transform", "improved feature matching", "improved feature coding" and "improved Gaussian mixture model" for image retrieval. They used Friedman test to compare alternatives for each framework differences. Eisinga et al. (2017) provided a computationally quick method to compute the exact p-value of the absolute rank sum difference of a pair of Friedman rank sums to be used in replacement of asymptotic results. Pereira et al. (2015) discussed many multiple comparison procedures for Friedman tests in case of population follows discrete distributions. They indicated that sign test is very conservative, and "Fisher's LSD" and "Tukey's HSD" tests computed with ranks are wider. Laurent and Turk (2013) discussed the test assumptions and misconceptions. Using simulation, they showed that the type I error can be larger than nominal value when the variance or skew of the treatment distributions differ. Gabor (2012) recommended that the results of the Friedman tests for marketing data can shape the base for applying "Kolmogorov-Smirnov test".

When the null hypothesis of Friedman test is rejected, the post-hoc pairwise comparisons using Nemenyi (1963) or Conover (1999) procedures are used to identify where the differences among treatments. As the number of groups increases, the number of required comparisons become large that leads to increase the type I error.

The aim of this paper is twofold. The main aim is to propose expression that uses in making a graphical approach for Friedman test by gathering the test and pairwise comparisons in one step. The second aim is to derive the sampling distribution of the proposed expression using method of moments as gamma distribution that helps in obtaining the decision limit. An application and simulation study are carried out to show the benefit of the method and to compute empirical type I error. The results are of great value for (a) huge reduction for the number of tests from $G(G-1)/2 + 1$ tests to $G$ tests, (b) providing visual, deep insight and understanding where the treatment effects occur.

The Friedman test is discussed in Section 2. The proposed method is introduced in Section 3. The empirical type I error is studied in Section 4. Section 5 is devoted for conclusion.

## 2  Friedman test

A randomized complete block design is a limited randomization plan in which the experimental individuals are first sorted into homogeneous blocks (row), and the treatments (groups) are allocated at random within the blocks; see, Friedman (1937), Eisinga et al. (2017) and Kandethody et al. (2021). The model for a randomized complete block design including the comparison of no interaction impacts, when both the row and group effects are fixed and there are $B$ blocks (BL) and $G$ groups (TR), is as

$$Y_{bg} = \theta + \rho_b + \tau_g + \varepsilon_{bg}$$

$\theta$ is a constant, $\rho_b$ are constants for the row (blocks) effects, $\tau_g$ are constants for the column (groups) effects and $\varepsilon_{bg}$ are independent errors, $b = 1,2,\dots,B$, $g = 1,2,\dots,G$.

Assume there are $G$ different groups (treatments) with values in every group $y_{bg}$, with block $b = 1,2,\dots,B$, and $n = BG$. Thus, the null hypothesis of equal treatment effects can be expressed as

$$H_0: \theta_1 = \theta_2 = \dots = \theta_G = \theta$$

versus at least two treatment effects are not equal; see, Conover (1990).

When analysing a randomized block design, sometimes the data consist of only the ranks within each block. Other times, it cannot be assumed that the data from each of the $g$ groups are from normally distributed populations. To conduct the test, the data values are replaced in each of the $B$ independent blocks with the corresponding ranks, so that the rank 1 is assigned to the smallest value in the block and rank $B$ to the largest. The test can be written as

$$F_g = \frac{12}{BG(G+1)} \sum_{g=1}^{G} R_g^2 - 3B(G+1)$$

Note that, $R_g$ is the sum for the ranks of $g$ treatment, $g = 1, 2, \ldots, G$.

The first three central moments of $F_g$ can be obtained from Gibbons and Chakraborti (2003) as

$$M_1 = E(F_g) = (G-1), \quad M_2 = V(F_g) = 2(G-1)\left(1 - \frac{1}{B}\right)$$

and

$$M_3 = 8(G-1)\left(1 - \frac{3}{B} + \frac{2}{B^2}\right)$$

The skewness can be computed as

$$sk = \frac{8(G-1)\left(1 - \frac{3}{B} + \frac{2}{B^2}\right)}{\left[2(G-1)\left(1 - \frac{1}{B}\right)\right]^{3/2}}$$

Therefore, As the number of blocks gets large, the test is approximated by using the chi-square distribution with $G - 1$ degrees of freedom as

$$F_g \sim \chi^2(G-1)$$

See; Sheskin (2000) and Gibbons and Chakraborti (2003).

In the case of Friedman test is significance, Nemenyi (1963) suggested post-hoc test for pairwise comparisons as

$$|\bar{R}_i - \bar{R}_j| > \frac{q_{\infty;G,\alpha}}{\sqrt{2}} \sqrt{\frac{G(G+1)}{6B}}$$

Note, $q_{\infty;G,\alpha}$ is the upper quantile of the studentized range distribution and this test is taking a family-wise error in consideration and is already conservative; see, Pohlert (2016).

Another post-hoc test for pairwise comparisons is proposed by Conover (1999) as

$$|\bar{R}_i - \bar{R}_j| > t_{1-\frac{\alpha}{2};(B-1)(G-1)} \sqrt{\frac{2G\left(1 - \frac{\chi_R^2}{B(G-1)}\right)\left(\sum_{i=1}^{B}\sum_{j=1}^{G} R_{ij}^2 - \frac{BG(G+1)^2}{4}\right)}{(G-1)(B-1)}}$$

Note, $t_{1-\frac{\alpha}{2};(B-1)(G-1)}$ is the upper quantile of the student t-distribution; see, Pohlert (2016).

## 3 Graphical approach for Friedman test

The Friedman test can be rewritten as

$$F_g = \frac{12}{BG(G+1)} \sum_{g=1}^{G} [R_g - 0.5B(G+1)]^2 = \sum_{g=1}^{G} \left[ \frac{R_g - 0.5B(G+1)}{\sqrt{\frac{BG(G+1)}{12}}} \right]^2$$

The contribution of every standardized group rank within $F_g$ is defined as

$$S_g = \left( \frac{R_g - 0.5B(G+1)}{\sqrt{\frac{BG(G+1)}{12}}} \right)^2, \quad g = 1, \ldots, G$$

Therefore, the Friedman's test could be plotted as

$$x_{axis} = g \text{ versus } y_{axis} = S_g \text{ with DL, for } g = 1,2,\ldots,G$$

Where DL is the decision limit obtained from the sampling distribution of $S_g$.

### 3.1 Approximation of sampling distribution of $S_g$

To investigate the approximation of the sampling distribution of $S_g$, the method of moments is used by equating the moment of $S_g$ with the moments of gamma distribution where the chi square distribution is a special case of gamma distribution. Since the groups are identically distributed, the first three moments of $S_g$ can be obtained from moments of $F_g$ as

$$M_1 = E(S_g) = 1 - \frac{1}{G}, \quad M_2 = V(S_g) = \left(2 - \frac{2}{G}\right)\left(1 - \frac{1}{B}\right)$$

and

$$M_3 = \left(8 - \frac{8}{G}\right)\left(1 - \frac{3}{B} + \frac{2}{B^2}\right)$$

The skewness can be obtained as

$$sk = \frac{\left(8 - \frac{8}{G}\right)\left(1 - \frac{3}{B} + \frac{2}{B^2}\right)}{\left[\left(2 - \frac{2}{G}\right)\left(1 - \frac{1}{B}\right)\right]^{3/2}}$$

The gamma distribution is

$$f(x) = \frac{\beta^\alpha}{\Gamma(\alpha)} x^{\alpha-1} e^{-\beta x}$$

Note that, $\beta$ is the scale, $\alpha$ is the shape and $x > 0$.

From Wackerly et al. (2018), the first four moments are

$$M_1 = E(X) = \frac{\alpha}{\beta}, \quad M_2 = V(X) = \frac{\alpha}{\beta^2}$$

and

$$sk = \frac{2}{\sqrt{\alpha}}, \quad Ku = \frac{6}{\alpha}$$

Since there are two parameters $\beta$ and $\alpha$, the two moments are required ($M_1$ and $M_2$ or $M_1$ and $sk$ or $M_2$ and $sk$). Because the $\alpha$ is shape parameter, it is preferable to use $M_1$ and $sk$.

By using skewness

$$\frac{2}{\sqrt{\alpha}} = \frac{\left(8 - \frac{8}{G}\right)\left(1 - \frac{3}{B} + \frac{2}{B^2}\right)}{\left[\left(2 - \frac{2}{G}\right)\left(1 - \frac{1}{B}\right)\right]^{3/2}}$$

Hence,

$$\hat{\alpha} = \frac{4\left[\left(2 - \frac{2}{G}\right)\left(1 - \frac{1}{B}\right)\right]^3}{\left[\left(8 - \frac{8}{G}\right)\left(1 - \frac{3}{B} + \frac{2}{B^2}\right)\right]^2}$$

By using mean

$$\frac{\alpha}{\beta} = 1 - \frac{1}{G}$$

Then

$$\hat{\beta} = \frac{\hat{\alpha}}{1 - 1/G} = \frac{4\left[\left(2 - \frac{2}{G}\right)\left(1 - \frac{1}{B}\right)\right]^3}{\left(1 - \frac{1}{G}\right)\left[\left(8 - \frac{8}{G}\right)\left(1 - \frac{3}{B} + \frac{2}{B^2}\right)\right]^2}$$

Therefore, the sampling distribution of $S_g$ can be approximated as

$$S_g \approx \text{gamma}\left(\text{shape} = \hat{\alpha}, \text{rate} = \hat{\beta}\right)$$

For large $B$, the sampling distribution of $S_g$ approaches chi square distribution with one degree of freedom.

## 3.2 Empirical moments

To investigate how well the gamma approximation for $S_g$, A simulation study is carried out to obtain the empirical first four moments of $S_g$ at $G = 3$ and $G = 5$ from symmetric distribution (normal) and asymmetric distribution (exponential) using different sample sizes. Five collections of blocks are inspected $B = 3, 5, 10, 15$ and $25$.

The following steps are used in simulation:

1. Choose the required design such as $B = 5$ and $n_g = 10$,
2. Simulate data from a selected distribution with similar mean and variance,
3. Rank the data inside each block,
4. Compute $S_g, g = 1, \ldots, G$ for each group,
5. Compute the first four moments for each $S_g, g = 1, \ldots, G$,
6. Repeat this R times and calculate the mean for every design.

Tables 1 and 2 give the mean, variance, skewness, and kurtosis for $S_g, G = 3, 5$ from normal, exponential distributions as well as the theoretical approximated values from gamma distribution. These tables illustrate that:

1. the gamma distribution gives a good approximation for mean and skewness of $S_g$ for all values of $G$ and $B$.
2. the gamma distribution does not give a good approximation for variance and kurtosis of $S_g$ for small values of $B$ especially at $B = 3$.
3. In terms of first four moments the gamma distribution gives a very good approximation of $S_g$ for $B \geq 5$,

Table 1 empirical four moments of $S_g$ using simulated data from normal and exponential distributions, $G = 3$ and 10000 replications.

| B | | Normal | | | | Exponential | | | | Theoretical Approx. | | | |
|---|---|---|---|---|---|---|---|---|---|---|---|---|---|
| | | Mean | Var | Skew | Kurt | Mean | Var | Skew | Kurt | Mean | Var | Skew | Kurt |
| 3 | | | | | | | | | | | | | |
| | $S_1$ | 0.658 | 0.645 | 1.689 | 5.279 | 0.671 | 0.652 | 1.663 | 5.174 | 0.667 | 0.222 | 1.41 | 3.00 |
| | $S_2$ | 0.659 | 0.651 | 1.685 | 5.248 | 0.679 | 0.687 | 1.652 | 5.169 | 0.667 | 0.222 | 1.41 | 3.00 |
| | $S_3$ | 0.663 | 0.681 | 1.697 | 5.166 | 0.661 | 0.661 | 1.686 | 5.124 | 0.667 | 0.222 | 1.41 | 3.00 |
| 5 | | | | | | | | | | | | | |
| | $S_1$ | 0.668 | 0.773 | 2.189 | 8.623 | 0.683 | 0.778 | 2.067 | 7.949 | 0.667 | 0.600 | 2.32 | 8.10 |
| | $S_2$ | 0.671 | 0.776 | 2.169 | 8.563 | 0.659 | 0.719 | 2.112 | 8.361 | 0.667 | 0.600 | 2.32 | 8.10 |
| | $S_3$ | 0.680 | 0.807 | 2.184 | 7.941 | 0.680 | 0.729 | 2.184 | 8.591 | 0.667 | 0.600 | 2.32 | 8.10 |
| 10 | | | | | | | | | | | | | |
| | $S_1$ | 0.668 | 0.856 | 2.577 | 11.752 | 0.647 | 0.812 | 2.516 | 11.649 | 0.667 | 0.948 | 2.92 | 12.8 |
| | $S_2$ | 0.663 | 0.831 | 2.522 | 11.655 | 0.651 | 0.780 | 2.411 | 10.730 | 0.667 | 0.948 | 2.92 | 12.8 |
| | $S_3$ | 0.650 | 0.801 | 2.520 | 11.807 | 0.660 | 0.813 | 2.475 | 11.248 | 0.667 | 0.948 | 2.92 | 12.8 |
| 15 | | | | | | | | | | | | | |
| | $S_1$ | 0.669 | 0.827 | 2.451 | 11.651 | 0.655 | 0.816 | 2.645 | 12.826 | 0.667 | 1.073 | 3.10 | 14.4 |
| | $S_2$ | 0.669 | 0.861 | 2.642 | 12.884 | 0.652 | 0.818 | 2.554 | 11.801 | 0.667 | 1.073 | 3.10 | 14.4 |
| | $S_3$ | 0.668 | 0.831 | 2.592 | 12.663 | 0.654 | 0.797 | 2.682 | 12.302 | 0.667 | 1.073 | 3.10 | 14.4 |
| 25 | | | | | | | | | | | | | |
| | $S_1$ | 0.673 | 0.892 | 2.553 | 13.711 | 0.667 | 0.824 | 2.820 | 12.655 | 0.667 | 1.17 | 3.25 | 15.87 |
| | $S_2$ | 0.668 | 0.903 | 2.833 | 12.121 | 0.663 | 0.862 | 2.694 | 13.311 | 0.667 | 1.17 | 3.25 | 15.87 |
| | $S_3$ | 0.687 | 0.901 | 2.669 | 13.490 | 0.663 | 0.897 | 2.596 | 12.469 | 0.667 | 1.17 | 3.25 | 15.87 |

Table 2 empirical four moments of $S_g$ using simulated data from normal and exponential distributions, $G = 5$ and 10000 replications.

| | | $G = 5$ | | | | | | | | | | |
|---|---|---|---|---|---|---|---|---|---|---|---|---|
| B | | Normal | | | | Exponential | | | | Theoretical approx.. | | | |
| | | Mean | Var | Skew | Kurt | Mean | Var | Skew | Kurt | Mean | Var | Skew | Kurt |
| 3 | | | | | | | | | | | | | |
| | $S_1$ | 0.810 | 1.033 | 1.849 | 6.662 | 0.800 | 1.013 | 1.869 | 6.462 | 0.80 | 0.27 | 1.29 | 2.5 |
| | $S_2$ | 0.803 | 1.046 | 1.817 | 6.197 | 0.797 | 1.002 | 1.884 | 6.553 | 0.80 | 0.27 | 1.29 | 2.5 |
| | $S_3$ | 0.793 | 0.982 | 1.837 | 6.338 | 0.822 | 1.058 | 1.830 | 6.212 | 0.80 | 0.27 | 1.29 | 2.5 |
| | $S_4$ | 0.812 | 1.050 | 1.890 | 6.502 | 0.797 | 0.986 | 1.860 | 6.456 | 0.80 | 0.27 | 1.29 | 2.5 |
| | $S_5$ | 0.819 | 1.028 | 1.799 | 6.134 | 0.805 | 1.013 | 1.857 | 6.410 | 0.80 | 0.27 | 1.29 | 2.5 |
| 5 | | | | | | | | | | | | | |
| | $S_1$ | 0.803 | 1.112 | 2.212 | 9.988 | 0.786 | 1.092 | 2.286 | 9.756 | 0.80 | 0.72 | 2.12 | 6.75 |
| | $S_2$ | 0.799 | 1.120 | 2.244 | 9.195 | 0.797 | 1.146 | 2.344 | 10.066 | 0.80 | 0.72 | 2.12 | 6.75 |
| | $S_3$ | 0.795 | 1.094 | 2.276 | 9.494 | 0.780 | 1.049 | 2.211 | 9.841 | 0.80 | 0.72 | 2.12 | 6.75 |
| | $S_4$ | 0.806 | 1.124 | 2.236 | 9.203 | 0.784 | 1.069 | 2.269 | 9.483 | 0.80 | 0.72 | 2.12 | 6.75 |
| | $S_5$ | 0.804 | 1.107 | 2.253 | 9.375 | 0.772 | 1.057 | 2.231 | 8.997 | 0.80 | 0.72 | 2.12 | 6.75 |
| 10 | | | | | | | | | | | | | |
| | $S_1$ | 0.803 | 1.154 | 2.327 | 10.123 | 0.784 | 1.165 | 2.560 | 12.122 | 0.80 | 1.14 | 2.66 | 10.66 |
| | $S_2$ | 0.822 | 1.307 | 2.633 | 12.719 | 0.797 | 1.174 | 2.413 | 10.695 | 0.80 | 1.14 | 2.66 | 10.66 |
| | $S_3$ | 0.817 | 1.271 | 2.640 | 13.270 | 0.786 | 1.151 | 2.509 | 11.459 | 0.80 | 1.14 | 2.66 | 10.66 |
| | $S_4$ | 0.790 | 1.178 | 2.541 | 11.781 | 0.787 | 1.122 | 2.445 | 11.186 | 0.80 | 1.14 | 2.66 | 10.66 |
| | $S_5$ | 0.797 | 1.221 | 2.601 | 12.394 | 0.792 | 1.121 | 2.345 | 11.094 | 0.80 | 1.14 | 2.66 | 10.66 |
| 15 | | | | | | | | | | | | | |
| | $S_1$ | 0.827 | 1.297 | 2.625 | 12.901 | 0.791 | 1.168 | 2.501 | 11.449 | 0.80 | 1.28 | 2.83 | 12.07 |
| | $S_2$ | 0.831 | 1.309 | 2.555 | 12.162 | 0.808 | 1.238 | 2.505 | 11.336 | 0.80 | 1.28 | 2.83 | 12.07 |
| | $S_3$ | 0.806 | 1.230 | 2.656 | 13.321 | 0.807 | 1.228 | 2.621 | 12.940 | 0.80 | 1.28 | 2.83 | 12.07 |
| | $S_4$ | 0.806 | 1.230 | 2.655 | 13.321 | 0.796 | 1.214 | 2.528 | 11.727 | 0.80 | 1.28 | 2.83 | 12.07 |
| | $S_5$ | 0.809 | 1.262 | 2.538 | 11.645 | 0.792 | 1.170 | 2.494 | 11.329 | 0.80 | 1.28 | 2.83 | 12.07 |
| 25 | | | | | | | | | | | | | |
| | $S_1$ | 0.790 | 1.174 | 2.686 | 14.462 | 0.795 | 1.264 | 2.690 | 12.826 | 0.80 | 1.41 | 2.96 | 13.22 |
| | $S_2$ | 0.802 | 1.297 | 2.781 | 14.634 | 0.812 | 1.288 | 2.760 | 14.548 | 0.80 | 1.41 | 2.96 | 13.22 |
| | $S_3$ | 0.815 | 1.249 | 2.631 | 12.877 | 0.777 | 1.137 | 2.782 | 14.808 | 0.80 | 1.41 | 2.96 | 13.22 |
| | $S_4$ | 0.778 | 1.202 | 2.980 | 15.199 | 0.801 | 1.244 | 2.747 | 14.319 | 0.80 | 1.41 | 2.96 | 13.22 |
| | $S_5$ | 0.783 | 1.193 | 2.784 | 14.640 | 0.815 | 1.316 | 2.734 | 13.397 | 0.80 | 1.41 | 2.96 | 13.22 |

Moreover, Figure 1 displays the histogram for $S_g, g = 1,2,3$ with gamma density function superimposed using simulated data from normal and exponential distributions using $G = 3, n_g = 15$. It can see that the gamma distribution gives a very good fit for $S_g$.

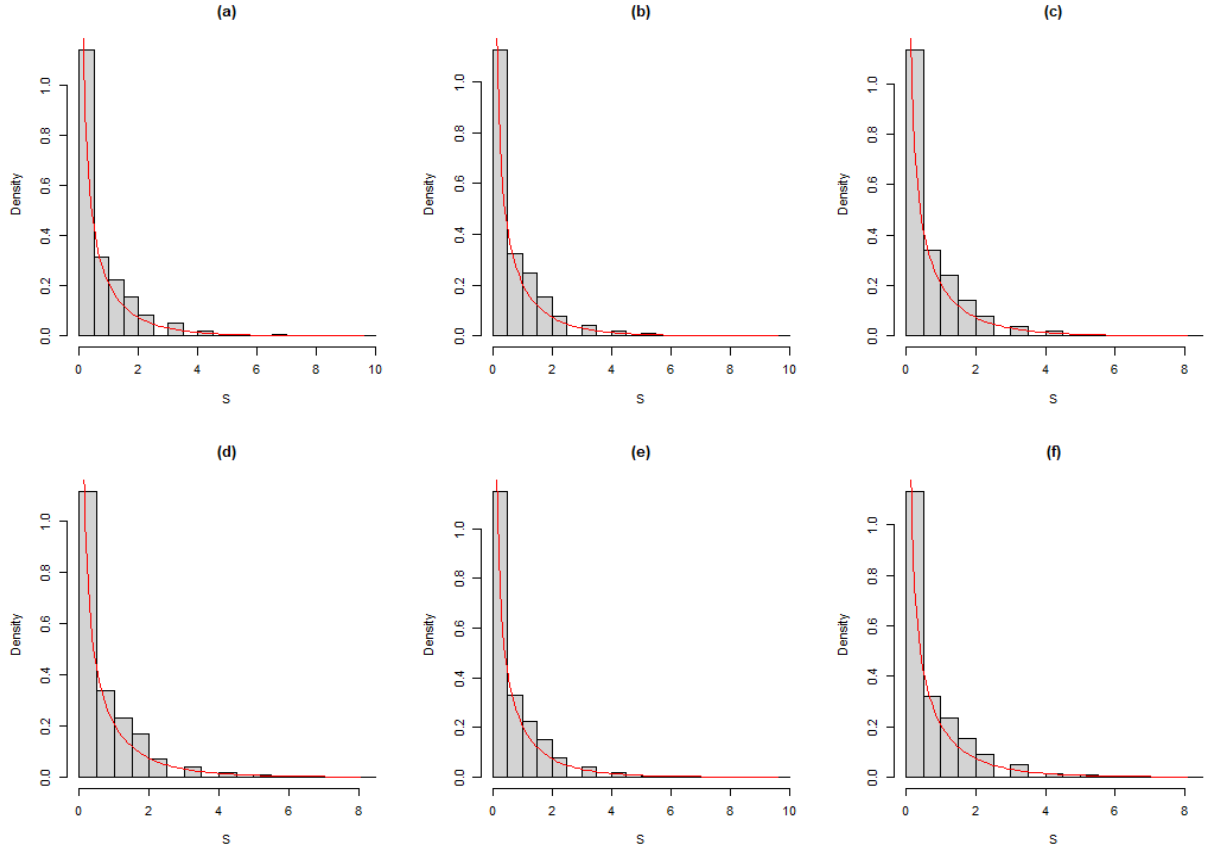

Figure 1 histogram for $S_g, g = 1,2,3$ with gamma density function superimposed using simulated data from normal distribution (a), (b) and (c) and exponential distribution (d), (e) and (f) using $G = 3, n_g = 15$ and 10000 the number of replications.

## 3.3 Decision limit and S plot

To obtain the decision line for $S_g, g = 1, ..., G$, it must consider $G$ tests and it is required to make difference between two kinds of $\alpha$ when performing many tests:
1. Type one error probability ("alpha per test") when working with a specific test. This is known as "test-wise alpha" ($\alpha[PT]$).
2. Type one error probability ("alpha per family of tests") for the whole "family of tests". This is known as "family-wise alpha" ($\alpha[PF]$) or "the experiment-wise alpha".

From Abdi (2007) the probability of committing type one error for a set of $G$ tests can be defined as

$$\alpha(PF) = 1 - (1 - \alpha(PT))^G$$

Therefore, alpha per test is

$$\alpha(PT) = 1 - (1 - \alpha(PF))^{1/G}$$

The Bonferroni approximation which is a simpler is

$$\alpha(PT) \approx \frac{\alpha(PF)}{G}$$

As an example, to perform $G = 5$, and the family-wise alpha $(PF) = 0.05$, by using the Bonferroni approximation, a test will be significance if its related probability is less than

$$\alpha(PT) \approx \frac{\alpha(PF)}{G} = \frac{0.05}{5} = 0.01$$

Therefore, the decision limit could be proposed by using the quantile function of gamma distribution and the Bonferroni approximation as

$$DL = \text{qgamma}\left(1 - \frac{\alpha}{G}, \text{shape} = \hat{\alpha}, \text{rate} = \hat{\beta}\right)$$

Hence,

$$\text{if any } S_g > DL, \text{ for } g = 1, 2, \ldots, G$$

$H_0$ is rejected.

The S-plot can be plotted as

$$x_{axis} = g \text{ versus } y_{axis} = S_g, \text{ with decision limit } DL$$

If any point outside the decision limit, $H_0$ is rejected and this will identify where the differences occur.

## 3.4 Application

This application from Gibbons and Chakraborti (2003, page 460) where number of questions asked by children in atmosphere like class with known person as a teacher. The 46 children are divided randomly into mutually exclusive groups of different sizes (A=24, B=12, C=6 and D=4). The questions asked by each one is recorded for 30 minutes on each of 8 days. The required is to test the hypothesis if there is no effect for the group size in terms of the total number of questions asked. The data for this application is shown in Table 3.

Table 3. The total number of questions asked on each of eight different days.

|     | Size |    |    |    |
| --- | --- | --- | --- | --- |
| Day | A | B | C | D |
| 1 | 14 | 23 | 26 | 30 |
| 2 | 19 | 25 | 24 | 33 |
| 3 | 17 | 22 | 29 | 28 |
| 4 | 17 | 21 | 28 | 27 |
| 5 | 16 | 24 | 28 | 32 |
| 6 | 15 | 23 | 27 | 36 |
| 7 | 18 | 26 | 27 | 25 |
| 8 | 16 | 22 | 30 | 32 |

The Friedman test, Nemenyi post-hoc, and Conover post-hoc tests are reported in Table 4. Note that the Friedman test and post-hoc are computed using R-software and MMCMR package in R; see, Pohlert (2016) and R Core Team (2021). Where the p-value for Friedman test is 0.0003, the null hypothesis of no effect among sizes is rejected. In addition, the Nemenyi post-hoc pairwise comparisons show that the differences are among groups A-C, A-D. This may give indication that the effect occurs because of group A. The Conover post-hoc pairwise comparisons show that the differences are among groups A-B, A-C, A-D, B-C and B-D. Note that the number of post-hoc tests needed is $G(G-1)/2 = 4(3)/2 = 6$ tests.

Table 4. Friedman test, Nemenyi post-hoc, and Conover post-hoc tests

|  | Friedman |  | Nemenyi post-hoc | | | Conover post-hoc | | |
| --- | --- | --- | --- | --- | --- | --- | --- | --- |
| Chi square | 18.6 |  | A | B | C | A | B | C |
| p-value | 0.0003 | B | 0.2127 | - | - | 0.0002 | - | - |
|  |  | C | 0.0027 | 0.4080 | - | 0 | 0.0019 | - |
|  |  | D | 0.0006 | 0.2127 | 0.9802 | 0 | 0.0002 | 0.3865 |

Figure 2 shows the S plot for the total number of questions asked on each of eight different days using 0.05 level of significance. Where the $S_g$ for groups A and D are outside the DL, the null hypothesis of no group size effect is rejected. In addition, the plot may directly indicate the effect occurs because of groups A and D. also, it can conclude that 58% of difference attributed to group A (10.8/18.6), followed by group D with 25.8% (4.8/18.6).

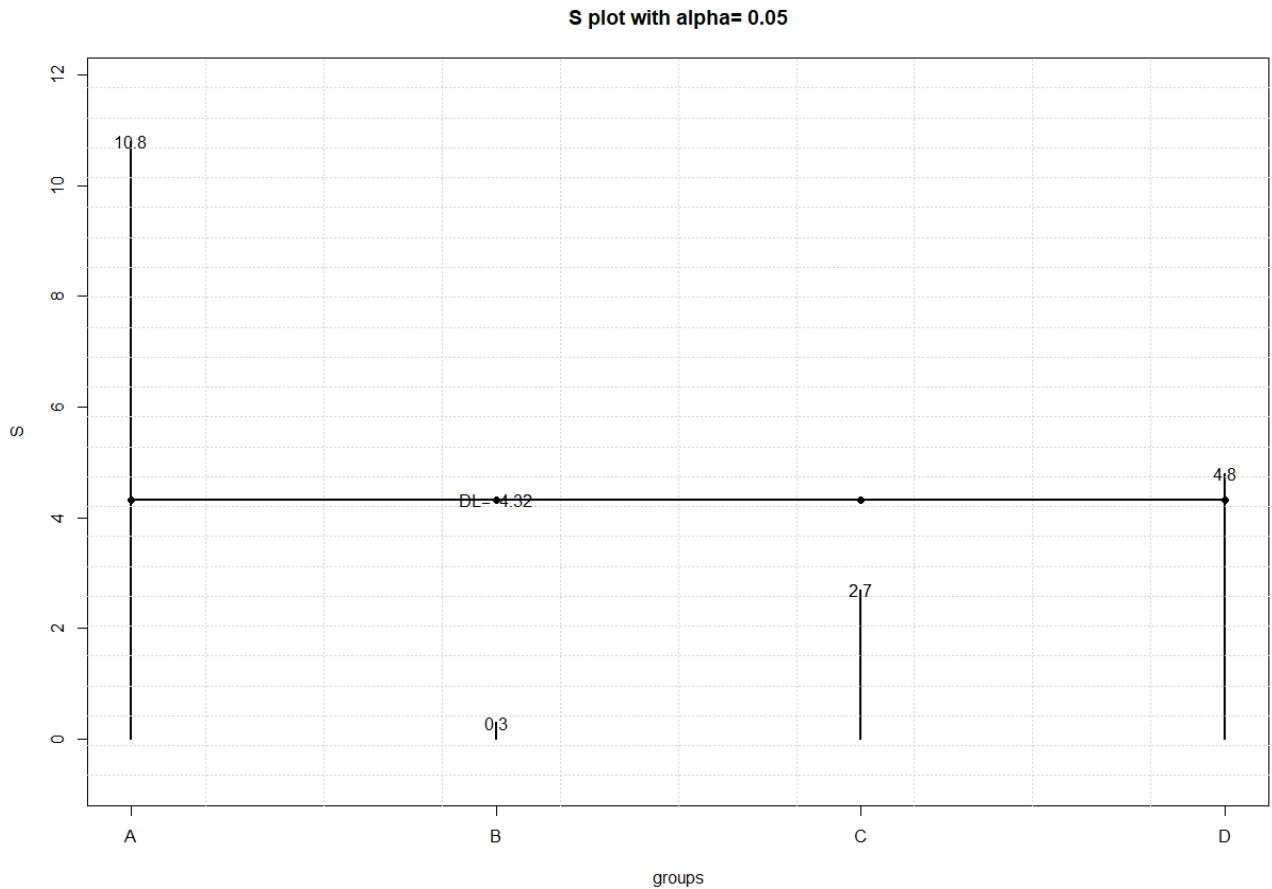

Figure 2. S plot for the total number of questions asked on each of eight different days using
$$\alpha[PT] = 0.0125$$

Figure 3 shows the S plot for the total number of questions asked on each of eight different days using 0.01 level of significance. Where the $S_g$ for group is outside the DL, the null hypothesis of no group size effect is rejected. Moreover, the plot may indicate the effect occurs because of group A.

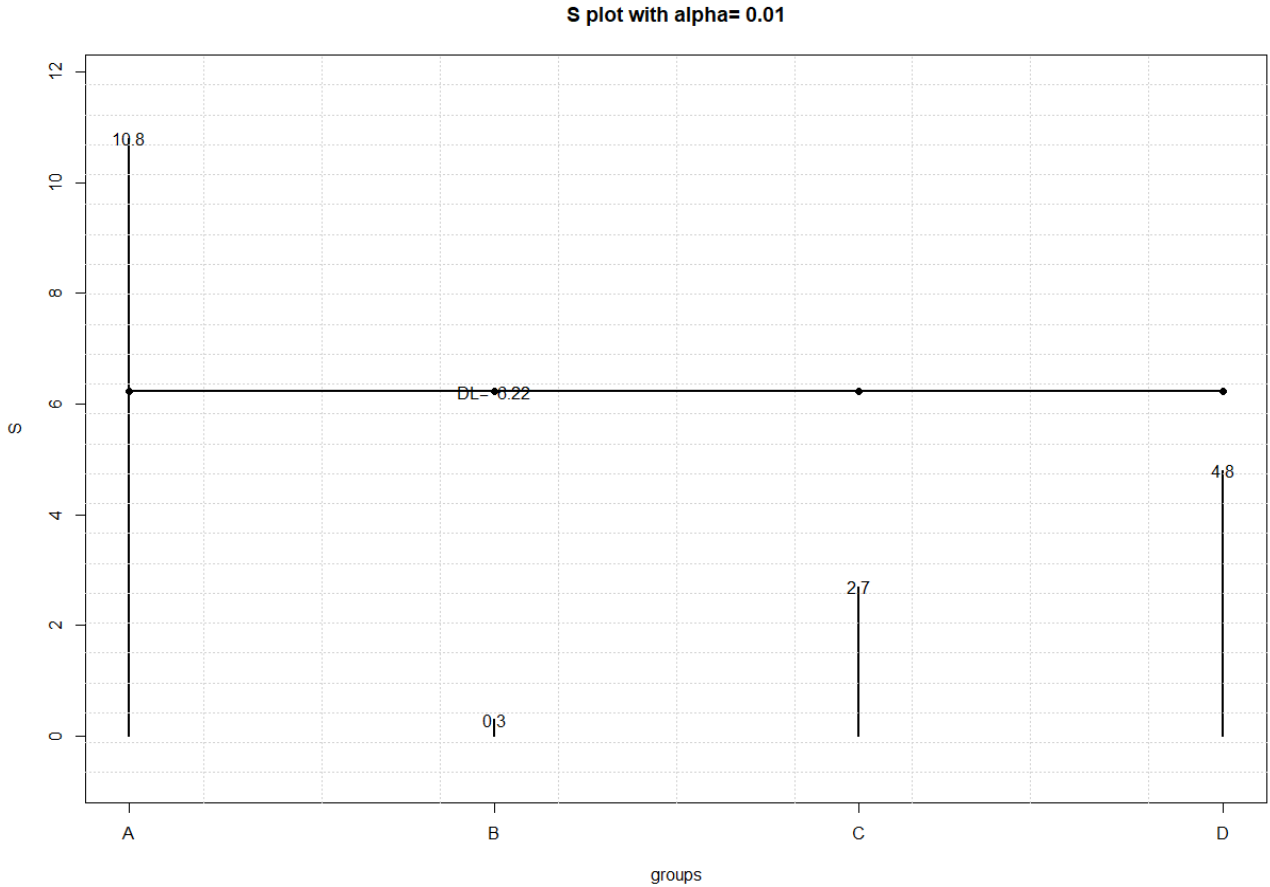

Figure 2. S plot for the total number of questions asked on each of eight different days using $\alpha[PT] = 0.0025$

## 4 Simulation study

### 4.1 Empirical type I error

The proposed method (S plot) using Bonferroni approximation is compared with Friedman test in terms of nominal one error (level of significance) at 0.05 and 0.01. Two settings were manipulated in the study: (a) group sizes $G$ (5, 10 and 20) and (b) the number of blocks B (8, 15, 25 and 50). To consider the effects of distributional shape on nominal one error, the normal and exponential distributions were chosen. To judge a particular design under which a test is non robust, the conditions of Bradley (1978) is employed. According to these conditions, a test is considered robust if its empirical rate of type one error $\alpha$ must be inside the interval $\alpha \pm \varepsilon$. Bradley selected $\varepsilon = 0.025$ which looks large. Therefore, the choice is something between 0 and 0.025, i.e., $\varepsilon = 0.015$. For 5% level of significance, a test considers robust in a particular case if its empirical type one error rate fall in the interval $0.035 \leq \hat{\alpha} \leq 0.065$ and for 1% level of significance, a test considers robust if its empirical type I error rate fall within the interval $0 \leq \hat{\alpha} \leq 0.02$. Correspondingly, a test considers non robust if its type one error rate does not fall in this interval. In any case, there is no common strategies by which tests are judged to be robust, so different explanations of the results are possible. Note that the Friedman test is computed using R-software, see, R Core Team (2021).

The following steps are used to estimate type I error ($p(\text{reject } H_0 | H_0 \text{ is true})$):

1. Construct the required design such as $\alpha = 0.05$, $B = 8$ and $n_g = 5$,
2. Simulate data from a desired distribution with the same mean for the required design,
3. Compute $F_r$ and $S_g, g = 1, \ldots, G$ for each group,
4. Find the decision limit (DL) for $S_g$ and p-value for Friedman test,
5. Make a dummy variable by giving 1 for reject and 0 else,
6. Repeat this R times and calculate the mean for every design.

Table 5 empirical type one error for $F_g$ and $S_g$ tests using Bonferroni approximation for $G = 5, 10$ and $20$, $B = 8, 15, 25$ and $50$, simulated data from Norm (normal) and Expo (exponential) distributions and 10000 replications.

|   | Norm. | Norm. | Norm. | Norm. | Expo. | Expo. | Expo. | Expo. |
|---|---|---|---|---|---|---|---|---|
| $\alpha$ | 0.05 | 0.05 | 0.01 | 0.01 | 0.05 | 0.05 | 0.01 | 0.01 |
|   | $F_g$ | $S_g$ | $F_g$ | $S_g$ | $F_g$ | $S_g$ | $F_g$ | $S_g$ |
| B | | | | $G = 5$ | | | | |
| 8 | 0.045 | 0.040 | 0.005 | 0.005 | 0.042 | 0.031 | 0.007 | 0.006 |
| 15 | 0.047 | 0.057 | 0.008 | 0.011 | 0.043 | 0.057 | 0.008 | 0.010 |
| 25 | 0.048 | 0.060 | 0.008 | 0.009 | 0.044 | 0.060 | 0.008 | 0.010 |
| 50 | 0.049 | 0.050 | 0.011 | 0.012 | 0.052 | 0.050 | 0.009 | 0.011 |
| | | | | $G = 10$ | | | | |
| 8 | 0.040 | 0.042 | 0.005 | 0.005 | 0.042 | 0.043 | 0.005 | 0.005 |
| 15 | 0.040 | 0.041 | 0.005 | 0.008 | 0.043 | 0.041 | 0.008 | 0.009 |
| 25 | 0.046 | 0.048 | 0.010 | 0.010 | 0.046 | 0.047 | 0.009 | 0.009 |
| 50 | 0.048 | 0.047 | 0.009 | 0.009 | 0.047 | 0.048 | 0.009 | 0.010 |
| | | | | $G = 20$ | | | | |
| 8 | 0.038 | 0.033 | 0.005 | 0.005 | 0.042 | 0.035 | 0.005 | 0.006 |
| 15 | 0.045 | 0.041 | 0.007 | 0.007 | 0.044 | 0.044 | 0.007 | 0.007 |
| 25 | 0.047 | 0.043 | 0.010 | 0.008 | 0.045 | 0.044 | 0.009 | 0.009 |
| 50 | 0.048 | 0.046 | 0.010 | 0.009 | 0.048 | 0.045 | 0.011 | 0.010 |

The simulation results for empirical type one error are given in Table 5. The results show that:

1. The type one error is robust for both methods $F_g$ and $S_g$ in terms of Bradley criteria,
2. The type one error is more sensitive for the number of blocks more than the number of groups,
3. As the numbers of blocks increase, the empirical type I approaches its nominal values,
4. The behaviour of empirical type I errors for $S_g$ is almost the same as $F_g$ for given $G$ and $B$.

These results confirm the robustness of the decision limit for proposed method in comparison with Friedman test in terms of Bradley criteria.

## 4.2 Discussion

It should be mentioned here that there are many alternatives to Bonferroni approximation such as "Holm's method" (1979), "Simes-Hochberg method" (Simes, 1986 and Hochberg, 1988), "Hommel's method" (Hommel, 1988 and 1989) and "Benjamini and Hochberg" (1995) method. All these methods are available in R-software under the function "p.adjust(p; method = ""; n = length(p))". These methods are c("holm", "hochberg", "hommel", "bonferroni", "BH", "BY"); BH: "Benjamini–Hochberg" and BY: "Benjamini and Yekutieli"; see, R Core Team (2021). The simulation study carried out using all these methods and all of them had given the same results with respect to type I error. Therefore, The Bonferroni results are only reported in Table 5.

With respect to ties in the data, there are many ways to treat ties in R-software where the function rank in R-software can be used; see, R Core Team (2021). This function takes the form:

rank(x, na.last = TRUE, ties.method = c("average", "first", "last", "random", "max", "min"))

one of the good methods that can be used is "random". This will treat the problem of ties in the data.; see, Kloke and Mckean (2020).

## 5 Conclusion

A graphical approach for Friedman test is proposed as the square of the standardized group rank in the randomized complete block design. The sum for all treatments is the Friedman test. The approximation of the sampling distribution for the proposed method had been derived as a gamma distribution based on method of moments. A decision limit was obtained based on the Bonferroni approximation and gamma distribution that permitted to showing Friedman test graphically. This graph called S plot that used to reject the null hypothesis if any value of $S_g$ falls outside the decision limit. The simulation results on type I error were shown robustness in terms of Bradley criteria for the proposed method.

This study had made many contributions (a) huge reduction in the number of required tests to illustrate where the differences occur from $G(G-1)/2 + 1$ tests to $G$ tests, (b) gathered the test and post-hoc comparisons in one step and kept comparative to classical test in terms of type I error, (c) provided visual representation where the treatment effects occurred, and (d) confirmed the robustness of the decision limit in terms of Bradley criteria.